\documentclass[a4paper,10pt]{article}

\usepackage{graphicx}
\usepackage{array}
\usepackage{multirow}

\usepackage[utf8]{inputenc}
\usepackage[T1]{fontenc}

\usepackage{geometry}
\usepackage{authblk}

\makeatletter
\renewcommand{\maketitle}{\bgroup
\begin{flushleft}
  \textbf{\Large\@title}

  \@author
\end{flushleft}\egroup
}
\makeatother

\usepackage{listings}
\usepackage{xcolor}
\title{Estimation of Basic Reproduction Number of the COVID-19 Epidemic in Denmark using a Two-Step Model}
\author[1]{Jan B. Valentin}

\affil[1]{Danish Center for Clinical Health Services Research (DACS), Department of Clinical Medicine, Aalborg University and Aalborg University Hospital, Aalborg, Denmark}

\date{}                     
\begin{document}

\maketitle

\section*{Abstract}
{\bf Objective:} To conduct an early estimation of the Basic Reproduction Number (BRN) induced by government interference, and to project resulting day to day number of in-patients, ICU-patients and cumulative number of deaths in a Danish setting.

{\bf Method:} We used the Kermack and McKendrick model with varying basic reproduction number to estimate number infected and age stratified percentages to estimate number of in-patients, ICU-patients and cumulative number of deaths. Changes in basic reproduction number was estimated based on current in-patient numbers.

{\bf Results:} The basic reproductive number in the time period of February 27th to March 18th was found to be 2.65, however, this number was reduced to 1.99 after March 18th.

{\bf Keywords:} COVID-19, basic reproduction number, Danish population

\section*{Introduction}
The COVID-19 virus have spread rapidly throughout the entire world and have already had a huge impact on the health care systems of most of the infected countries\cite{ferg2020,Verity2020.03.09.20033357}. As of March 20th 2020 the COVID-19 epidemic in Denmark have killed 34, while more than 80 people have been admitted to the intensive care units (ICU) with the first case confirmed by February 27th 2020\cite{ssd20mar}.

Learning from other countries the Danish government have imposed new legislation as well as other means of interference\cite{politi20mar} with the purpose of avoiding catastrophic numbers of deaths and total collapse of the health care system. Actions to reduce stress of the health care system includes; home isolation of confirmed cases, closing of schools, non-essential businesses and public workplaces, closing of country borders and restriction on gatherings of more than 10 people. These actions may help to create a much-needed delay of the peak number of infected, in order to establish the necessary resources to overcome the COVID-19 epidemic with fewest adverse outcomes.

The aim of the current study is to conduct an early estimation of the Basic Reproduction Number (BRN) induced by government interference, and to project resulting day to day number of in-patients, ICU-patients and cumulative number of deaths in a Danish setting.

\section*{Method}
\subsection*{Model}
We implemented the Kermack and McKendrick model\cite{kermack1997} in a two-step manner adjusting the BRN during the study period. This model was restricted to modelling number of infected, number of recovered and number of susceptible. The outbreak origin was assumed to be 14 days prior to first confirmed case in Denmark, thus origin was set at February 13th. At origin, we assumed one person to be infected and the remaining Danish population to be susceptible. The infectious time frame was assumed to be 6.5 days\cite{ferg2020}.
\subsection*{Data}
The Danish National Health Authority provided observed numbers of in-patients, ICU patients and cumulative number of deaths on national level\cite{ssd20mar}. Age distribution of the current Danish population as of January 2020 was obtained from Statistics Denmark\cite{st20mar}.
Data was available from March 16th until March 25th, both included, except for death counts, which were only available until March 24th.
Age-stratified and overall percentages of symptomatic, in-patient and ICU cases as well as mortality following COVID-19 infection were extracted from a recent report by Ferguson \emph{et al}\cite{ferg2020}.
\subsection*{Estimating number of excess in-patient beds needed}
Number of excess in-patient beds needed was estimated from age-stratified percentages of newly infected subjects, with an assumed 5 day delay accounting for a non-symptomatic infection period. This delay was implicit in the model. In addition, we assumed that 50\% percent of infected were non-symptomatic, thus, multiplying all age-stratified percentages by 0.5. Number of ICU beds needed was estimated from age-stratified percentages of number of in-patients. All in-patients including ICU cases were assumed to be hospitalised for 10 days.
\subsection*{Estimating cumulative number of deaths}
Cumulative number of deaths was estimated from age-stratified percentages of recovered subjects. A delay of 15 days was implemented to account for 5 days non-symptomatic infection period and 10 days hospitalisation. Finally, the infection period was subtracted.
\subsection*{Parametrization and optimization}
With the assumptions stated above, we were able to construct a three-parameter model. These parameters included; BRN in initial step (BRN$\!_1$), BRN in second step (BRN$\!_2$) and time of switch. The model was fitted to number of in-patient beds needed. We used a grid search approach on a 16-core Xeon(R) CPU E5-2630 v3 @ 2.40GHz hyperthreaded machinery, to find the optimal parameters measured by mean absolute error.
To investigate whether the model develops as more data is added, we fitted three models using data available up until March the 23rd, 24th and 25th, respectively. Once models was fitted, we projected number of in-patient and ICU beds needed as well as cumulative number of deaths. Projections were only presented for the model fitted on all available data. All projections was compared to the observed data.
All analyses were conducted using Python version 2.7 (Python Software Foundation. Python Language Reference, version 2.7. Available at www.python.org).

\section*{Results}
The BRN in the initial step was estimated at 2.65 and 1.99 in the second step. Time of switch was estimated to be March 18th 2020. All three models were optimized at these values, however, the mean absolute error ranged from 8.3 to 9.4 in-patient beds, where the final model had the highest mean absolute error.

\begin{figure}[hb!]
\begin{center}
\centerline{
\includegraphics[width=0.5\textwidth,trim=50 70 50 50,clip]{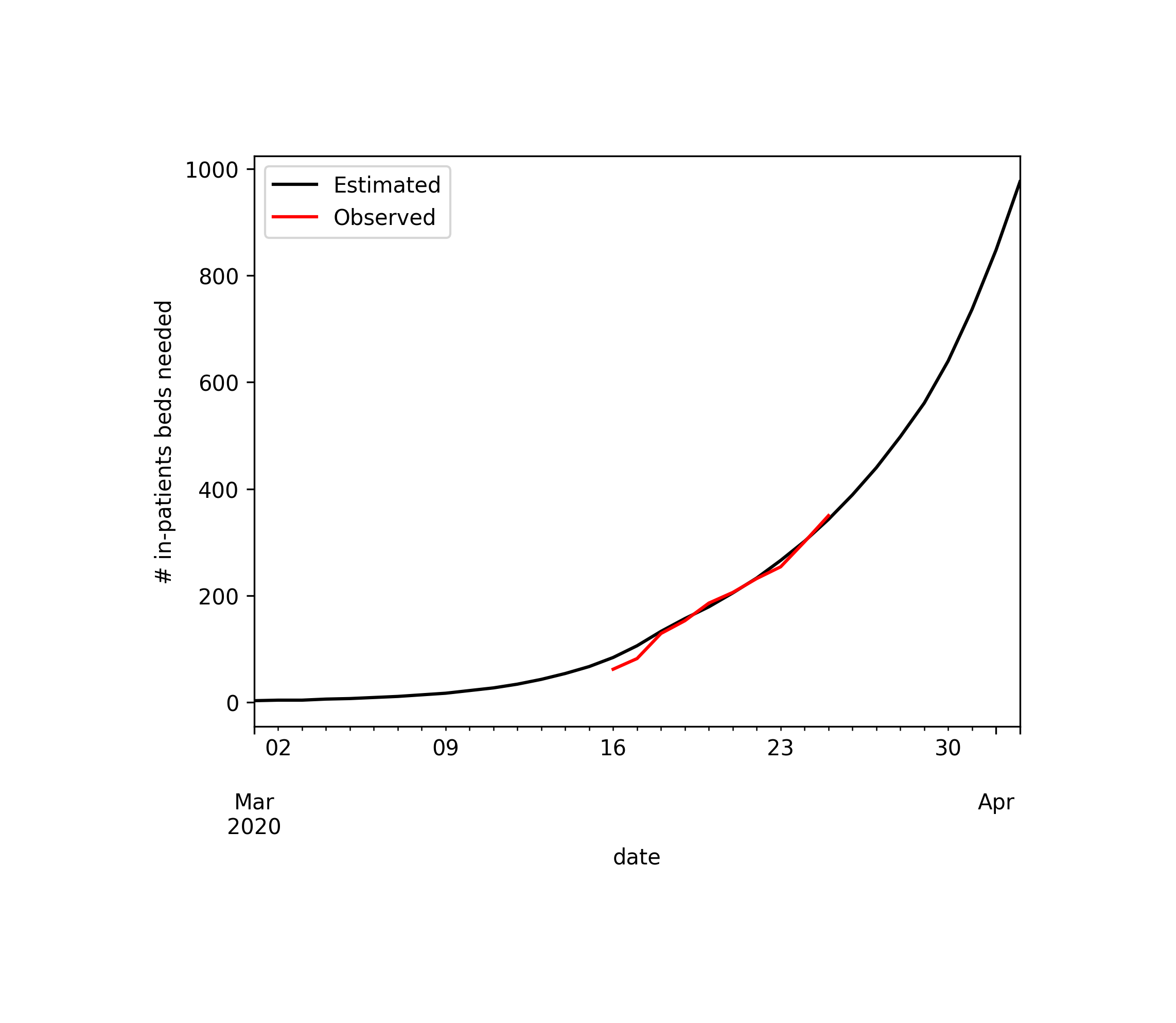}
\includegraphics[width=0.5\textwidth,trim=50 70 50 50,clip]{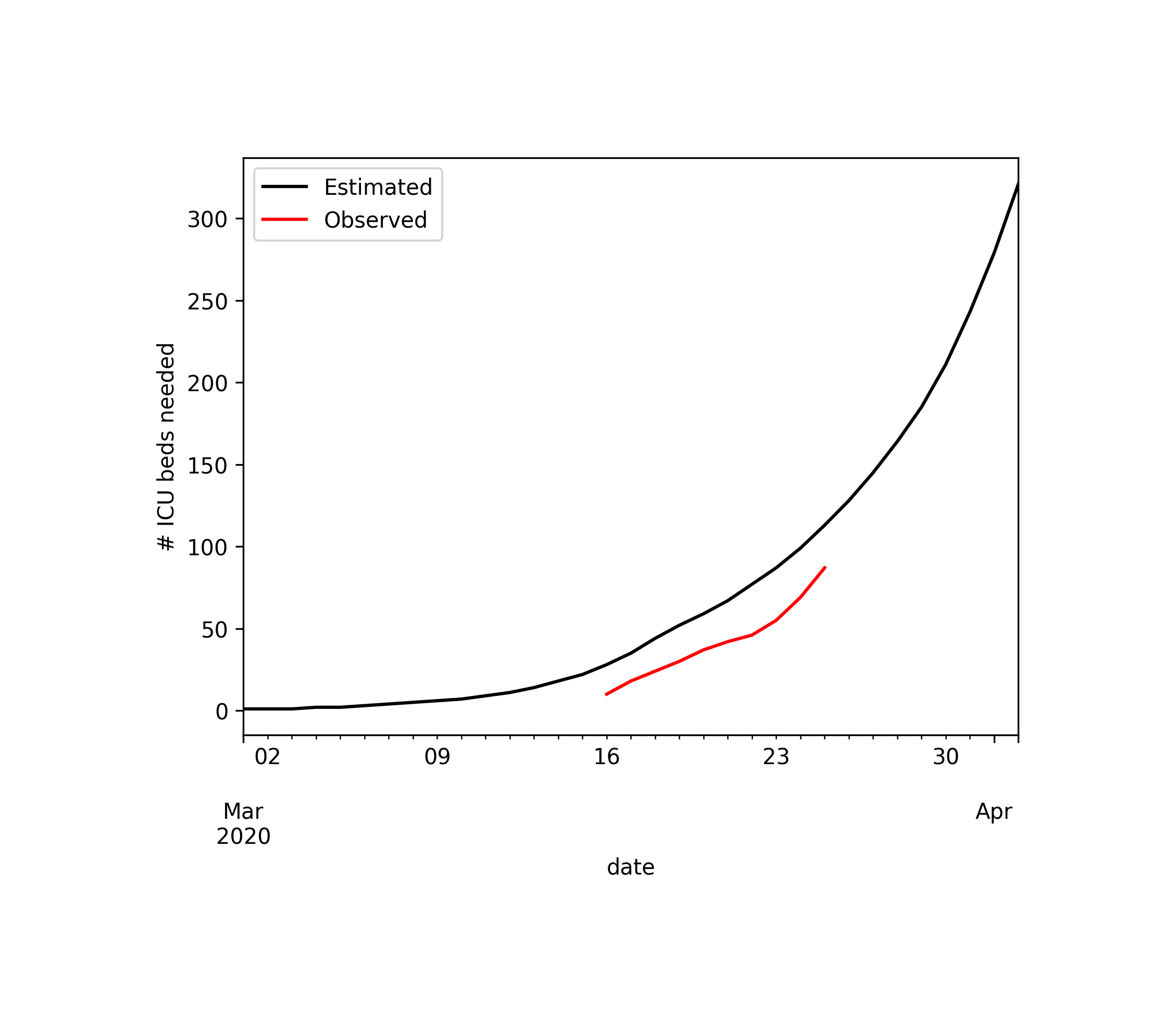}}
\end{center}
\caption{\footnotesize{\emph{Projected number of in-patient (left) and ICU (right) beds needed following COVID-19 infection along with observed data in the month of March.}}}
\label{fig:inpatients}
\end{figure}

\begin{figure}[hb!]
\begin{center}
\centerline{
\includegraphics[width=0.5\textwidth,trim=50 70 50 50,clip]{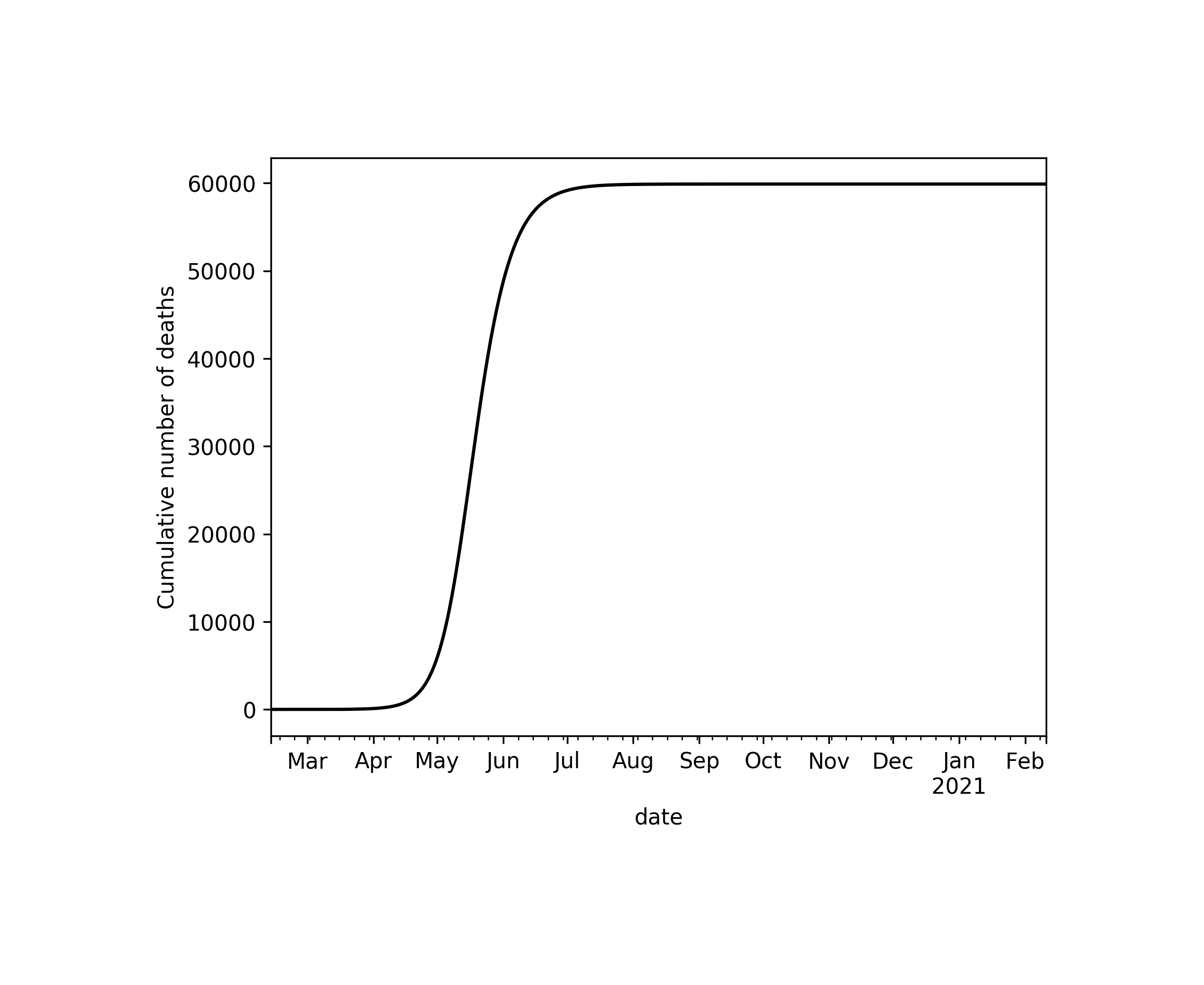}
\includegraphics[width=0.5\textwidth,trim=50 70 50 50,clip]{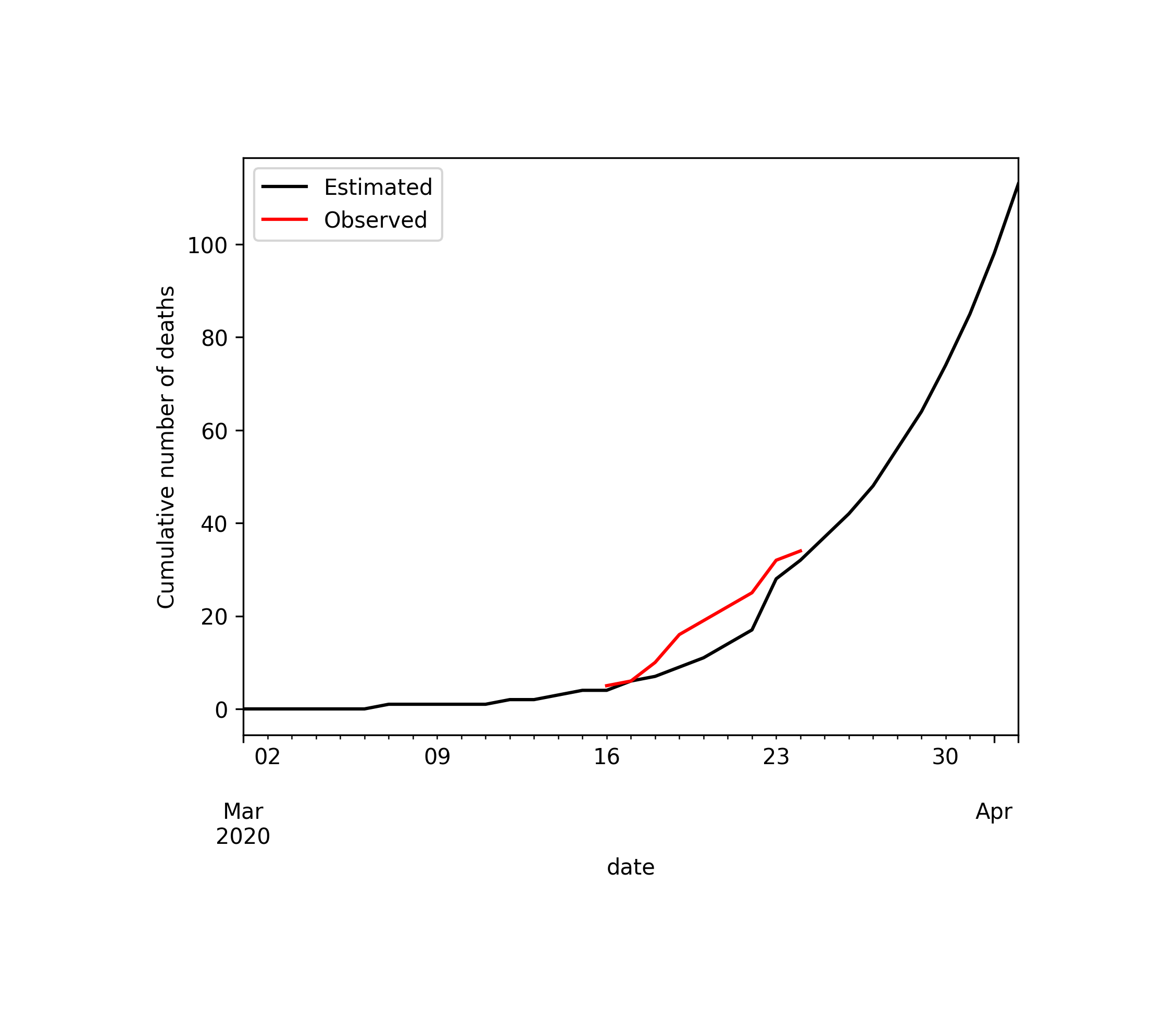}}
\end{center}
\caption{\footnotesize{\emph{Projected cumulated number of deaths caused following COVID-19 infection. Left: One year projection from origin. Right: Projection and observed data in the month of March.}}}
\label{fig:Dead}
\end{figure}

\begin{figure}[hb!]
\begin{center}
\centerline{
\includegraphics[width=0.5\textwidth,trim=50 70 50 50,clip]{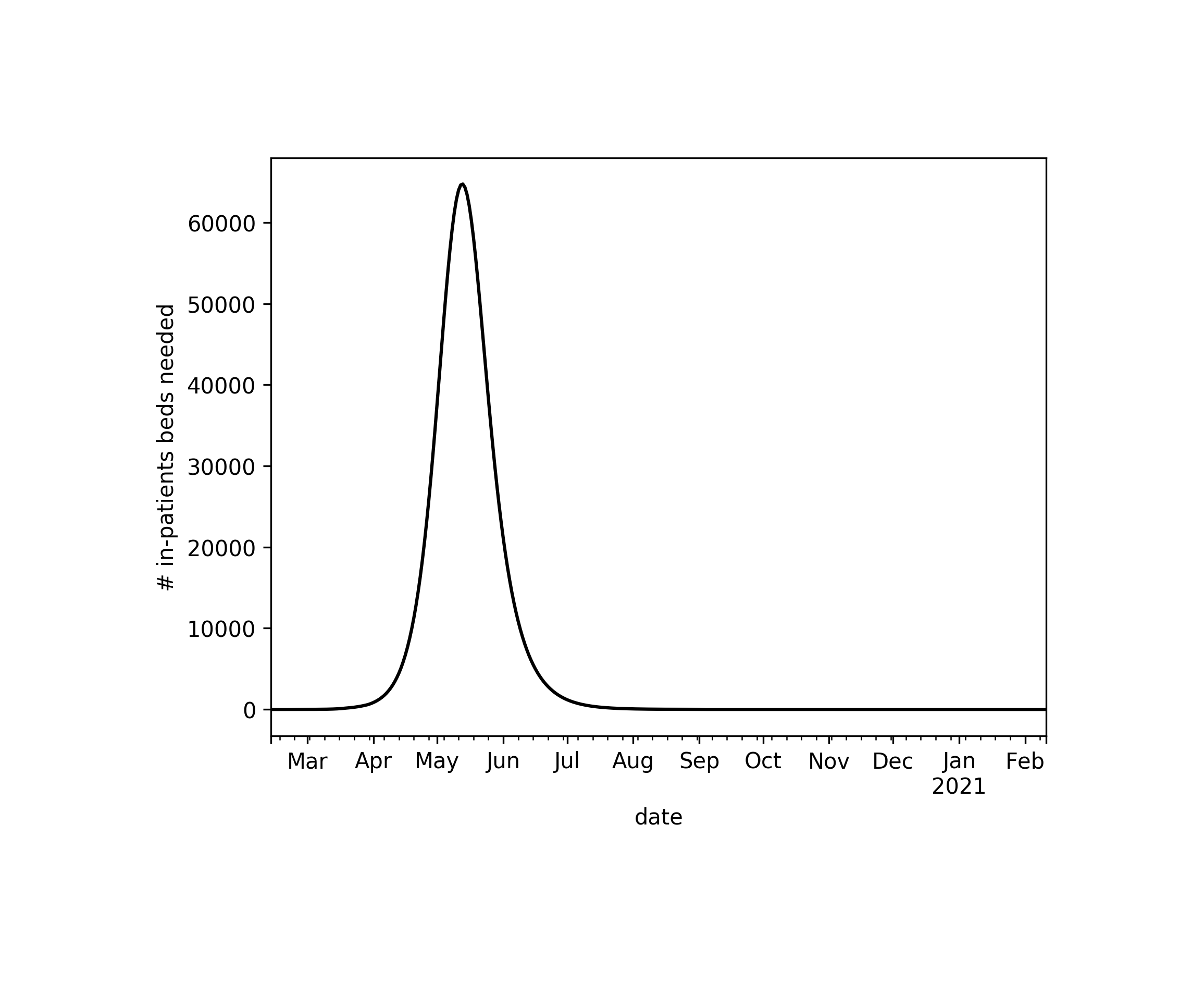}
\includegraphics[width=0.5\textwidth,trim=50 70 50 50,clip]{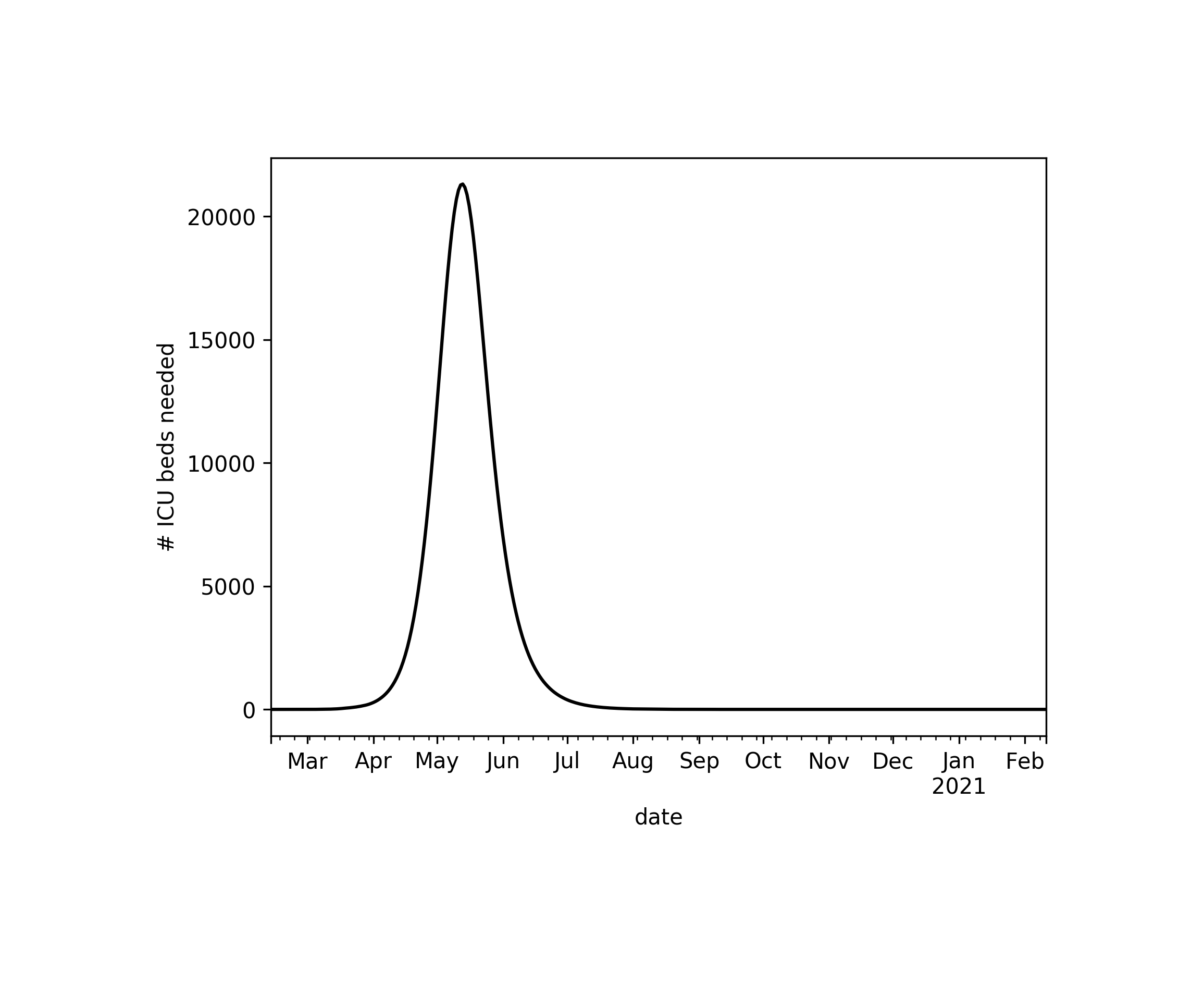}}
\end{center}
\caption{\footnotesize{\emph{Projected number of in-patient (left) and ICU (right) beds needed following COVID-19 infection, one year from origin.}}}
\label{fig:inpatientslong}
\end{figure}

\begin{figure}[hb!]
\begin{center}
\centerline{
\includegraphics[width=0.5\textwidth,trim=50 70 50 50,clip]{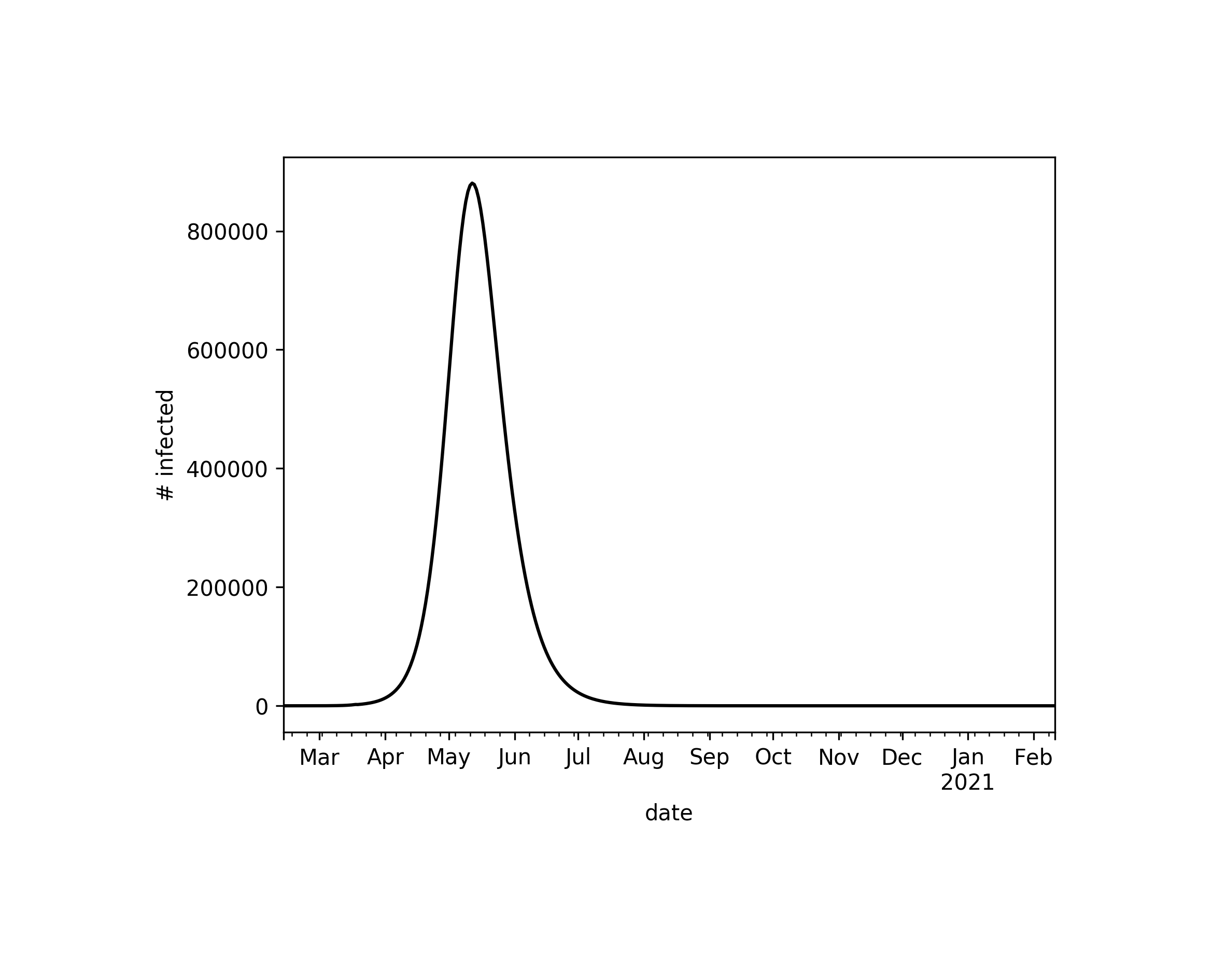}
\includegraphics[width=0.5\textwidth,trim=50 70 50 50,clip]{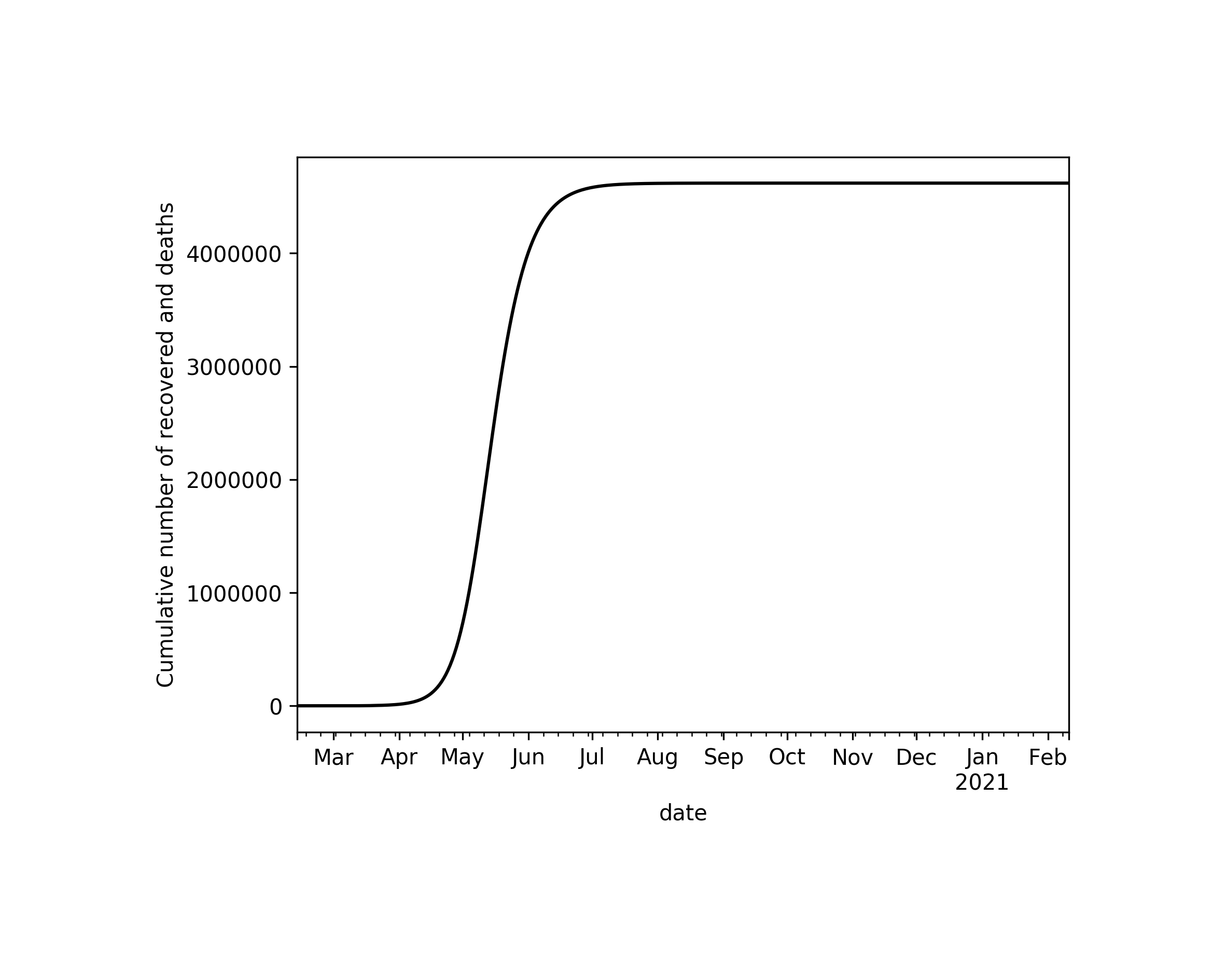}}
\end{center}
\caption{\footnotesize{\emph{Projected number of infected with (left) and recovered from (right) COVID-19, one year from origin. Recovered includes diseased.}}}
\label{fig:SIR}
\end{figure}

Figure \ref{fig:inpatients} shows the projections of day to day in-patient and ICU beds needed in the month of March, along with observed data. Figure \ref{fig:Dead} shows cumulative number of deaths following COVID-19 infection for the month of March, along with observed data. One year projection from origin is also shown in this figure. Figure \ref{fig:inpatientslong} shows the one year projections from origin of day to day in-patient and ICU beds needed. Figure \ref{fig:SIR} shows the one year projections from origin of day to day number of infected as well as recovered. Number of recovered includes diseased.

The timeline of the epidemic is depicted in the following along with a few key dates as well as the estimated time of switch from step 1 to step 2. The estimated time of switch occurred same day as last government directions were at effect.

\begin{tabular}{ >{\footnotesize}p{16mm}   >{\footnotesize}>{\raggedleft}p{54mm} | >{\footnotesize}p{54mm} }
\footnotesize
&&\\
&{\bf February 13th:} Model origin&\\
&&\\
&{\bf February 27th:} First case confirmed&{\bf February 27th:} Case isolation at home\\
&&\\
&&{\bf March 12th:} Planned lock-down of schools and non-essential public workplaces, everyone else are encourage to stay home\\
&&\\
&&{\bf March 14th:} Country borders are closed\\
&&\\
&{\bf March 16th:} Initial reports on number of in-patients available on region level&{\bf March 16th:} Previous orders are at effect\\
&&\\
&&{\bf March 17th:} Small non-essential businesses are ordered closed, gatherings of more than 10 people are ordered disbanded\\
&&\\
&&{\bf March 18th:} Previous orders are at effect, fines are imposed\\
BRN$=2.65$&&\\
\hline
BRN$=1.99$&&\\
&&{\bf March 19th}\\
&&\\
&{\bf March 25th:} Last available observation&\\
&&\\
\end{tabular}

\subsection*{Discussion}
The estimate of BRN$\!_1$ of 2.65 is in good correspondence with previous studies of COVID-19\cite{ferg2020}. However, this number largely depends on the assumption on the time of outbreak. Knowing that the COVID-19 virus is likely to have immigrated to Denmark with multiple hosts, the BRN in the initial step should be interpreted with some caution. The main purpose of the initial step is to set the level of infected in the second step. In the second step the BRN was estimated at 1.99, which is a reduction of 25\%. Step two is to a larger extend a reflection of the data, and thus not surprising the time of switch occurred simultaneously with a large-scale political interference. It is possible, however, that the full impact of the political interference is delayed as we may expect a gradual change of BRN rather than an abrupt switch, thus, more than two steps might be required as more data becomes available. For this reason we would also expect, that long-term projections based on the presented model will overshoot the actual counts.

\subsection*{Model limitations}
Besides the lack of individual-level data, the model is based on multiple assumptions, of which the assumptions on transmission generation time as well as age-stratified percentages requiring admission, requiring intensive care and mortality are of lesser concern. However, the assumption on 10 days of hospital stay are subject to some scrutiny. Hence, a change in the length of hospital stay would likely impact the estimation of the BRN in both steps. 
The assumption associated with the most doubt is the assumption that 50\% of cases are asymptomatic. The assumption is based on experience from other countries\cite{ferg2020,Verity2020.03.09.20033357}, but is likely obscured by large number of asymptomatic cases, which are never found. Moreover, in the current model we assumed that no one in the entire population was immune to COVID-19 at origin. A sample estimation of how many people of the Danish population are immune, at given time points would greatly improve the model.

\bibliographystyle{unsrt}
\bibliography{citations}

\end{document}